\documentstyle[12pt,axodraw]{article}
\newcommand{\fig}[1]{\item   \label{#1} }

\renewcommand{\baselinestretch}{1.0}
\renewcommand{\theequation}{\thesection.\arabic{equation}}
\renewcommand{\vec}[1]{{\bf #1}}

\newcommand{\btt}{\begin{tt}}
\newcommand{\ett}{\end{tt}}
\newcommand{\bsf}{\begin{it}}
\newcommand{\esf}{\end{it}}
\newcommand{\cl}{{\rm cl}}
\newcommand{\ini}{{\rm in}}
\newcommand{\nn}{\nonumber}
\newcommand{\mref}[1]{(\ref{#1})}

\newcommand{\prd}[3]{ {\it  Phys.\ Rev.\ } {\bf D{#1}} ({#2}) {#3}}
\newcommand{\prl}[3]{ {\it  Phys.\ Rev.\ Lett.\ } {\bf {#1}} ({#2}) {#3}}

\newcommand{\pl}[3]{ {\it Phys.\ Lett.\ }{\bf B{#1}} ({#2}) {#3}}
\begin{document}
\title{
	\begin{flushright}
	{\normalsize TPI--MINN--95--08/T \\
	NUC--MINN--95--11/T \\
	HEP--UMN--TH--1337 \\
	April 1995 \\}
	\end{flushright}
\bf
Really Computing Non-perturbative Real Time Correlation Functions }
\author{
	Dietrich B\"odeker and  Larry McLerran   \\
	{\small\it School of Physics and Astronomy,
	University of Minnesota, Minneapolis, MN 55455, USA} \\
	and\\
Andrei Smilga \\
{\small\it ITEP, B.Cheremushkinskaya 25, Moscow 117259, Russia}\\
 }

\date{}

\renewcommand{\baselinestretch}{1.5} 
\parindent=20pt

\maketitle

\begin{center}
{\bf Abstract}\\
\end{center}
It has been argued by Grigoriev and Rubakov that one can simulate real
time processes involving baryon number non-conservation at high
temperature using real time evolution of classical equations, and
summing over initial conditions with a classical thermal weight.  It
is known that such a naive algorithm is plagued by ultraviolet
divergences.  In quantum theory the divergences are regularized, but
the corresponding graphs involve the contributions from the hard
momentum region and also the new scale $\sim gT$ comes into play.  We
propose a modified algorithm which involves solving the classical
equations of motion for the effective hard thermal loop Hamiltonian
with an ultraviolet cutoff $\mu \gg gT$  and integrating over initial
conditions with a proper thermal weight.  Such an algorithm should
provide a determination of the infrared behavior of real time
correlation function $<Q(t) Q(0)>_T$ determining the baryon violation
rate. Hopefully, the results obtained in this modified algorithm would
be cutoff-independent.

\vfill \eject

\section{Introduction}

Sphaleron processes in electroweak theory mediate baryon number
violating processes, and have been computed to be significant at high
temperatures $\cite{manton} - \cite{arnold}$.  This computation
involves estimating the probability of a classical thermal fluctuation
evolving over the top of an energy barrier which separates
topologically distinct phases of electroweak theory.  At very high
temperatures, where the masses of the W and Z boson can be ignored,
the field configurations which dominate the contribution have a
typical size of order $1/\alpha_W T$ \cite{arnold} .  This is the
length scale on which non-perturbative physics is important, and there
is no known way of doing a weak coupling estimate of the effect.

In a provocative paper by Grigoriev and Rubakov $\cite{grigoriev}$, a
numerical algorithm was suggested for the estimation of the baryon
number violation rate.  They argue that at high temperature, the
dominant process for sphaleron processes involves transitions of
fields over the energy barrier. The dominant transitions should
therefore be classically allowed thermal processes.  Since the field
configurations useful for sphaleron processes are long wavelength and
involve many quanta, it is reasonable to assume that the evolution of
the field configuration over the top of the barrier is in fact
classical.

If the evolution of the fields is classical, then they will be
described by classical equations of motion with given initial
conditions.  These initial conditions are determined by a classical
thermal ensemble where the initial distribution of fields and their
conjugate momenta are computed in a Monte-Carlo simulation.

This method was used to compute the rate of sphaleron transitions in
electroweak theory at temperature higher than the electroweak phase
transition temperature $\cite{ambjorn}$.  It was also tested in two
dimensional models $\cite{shap} - \cite{bochkarev}$.  Although the two
dimensional results looked convincing, there have remained a number of
plaguing problems associated with obtaining correct dependences on
effective temperature dependent masses, and these masses depend on the
ultraviolet region of the theory where, as we will see, there are
problems associated with treating the theory classically.  The 3+1
dimensional simulations of electroweak theory are plagued by
ultraviolet singularities associated with making the spatial lattice
spacing smaller, and one may worry whether any quantitative numerical
conclusions are warranted.

It is easy to understand why there are problems associated with taking
the zero lattice spacing limit in such simulations.  In the second
section, we shall present a detailed mathematical argument in the
context of $\phi^4$ theory .  In the remainder of the Introduction, we
shall present heuristic arguments which display the nature of the
problem.

It has been well known since Rayleigh, Jeans, and Planck that, if one
tries to describe a bath of photons classically, the free energy is
ultraviolet divergent and depends on the cutoff.  To see this,
consider the energy density for a gas of scalar massless bosons.  This
density at high temperatures behaves as
\begin{eqnarray}
\label{Equ}
E = \int \frac{d^3p}{(2\pi)^3} \frac{|{\bf p}|}{e^{\beta |{\bf p}|} - 1}
= \frac{\pi^2 T^4}{30}
\end{eqnarray}
On the other hand, if we try to compute the thermodynamic
characteristics of the system classically, we have the functional
integral representation
\begin{eqnarray}
\label{Zcl}
	Z^{cl} = \int [d\phi] [d\Pi]~ exp\{-\beta H \}
\end{eqnarray}
where the Hamiltonian is
\begin{eqnarray}
	H = \int d^3x \{ {1 \over 2} \Pi (x)^2 + {1 \over 2} (\nabla \phi
(x) )^2 \}
\end{eqnarray}

The results are ultraviolet divergent. For the energy density we get
\begin{eqnarray}
\label{Ecl}
	E^{cl}(T) = \frac T{6\pi^2} \mu^3
\end{eqnarray}
where $\mu $ is a momentum ultraviolet cutoff (The simplest way to get
this result is to substitute the Bose distribution function in
Eq.(\ref{Equ}) by its classical limit $T/|{\bf p}|$~).  This is the
well known classical Rayleigh-Jeans divergence and was one of the
original motivations for quantum mechanics.

Ultraviolet divergences also crop up in other quantities. Consider
e.g. the correlator $<\phi(x,t)~\phi(0)>_T$ in $\phi^4$ theory.

This bears on what we discuss here since the quantity of interest
defining baryon nonconservation rate in the standard model is a
similar ( but, of course, more complicated) correlator
\begin{eqnarray}
   \label{QQ}
C(t) = <[Q(t) - Q(0)] Q(0)>_T
\end{eqnarray}
where
\begin{eqnarray}
\label{Chern}
Q(t) = \frac {g^2}{32\pi^2} \epsilon_{ijk} \int d^3x
\left( F^a_{ij} A^a_k - \frac g3 \epsilon^{abc} A^a_i A^b_j A^c_k \right)
\end{eqnarray}
is the Chern-Simons winding number of the $SU(2)$ gauge field
configuration.  The expectation is that the correlator (\ref{QQ})
behaves as $C(t) \sim \Gamma V |t|$ at large $|t| \gg (\alpha_W
T)^{-1}$ where $\Gamma$ is the baryon number nonconservation rate per
unit volume and is estimated as \cite{arnold}
\begin{eqnarray}
\label{DelB}
\Gamma^{\Delta B \neq 0} (T) = \kappa (\alpha_W T)^4
\end{eqnarray}

 We take the interaction term in the action as
\begin{eqnarray}
	S = \int dt \int d^3x ~ {\lambda \over {4! }} \phi^4 (x)
\end{eqnarray}
This yields in lowest order, the diagram shown in Figure
\ref{scalarloop}.

In this paper, all perturbative calculations are performed
consistently in the real time formalism which is the most suitable
tool to study kinetic properties in thermal medium. In its complete
form the formalism has been developed in \cite{indus,Keldysh}.  It
involves some intricacies, one should carefully distinguish the
retarded, the advanced and the so called P--component of the Green's
function
\footnote{The review can be found in \cite{LP,Landsman}. For
applications of the formalism to hot gauge theories see e.g.
\cite{Lebedev,damping} .}.  But we do not really need to go into
details here because, for our purposes (we will be interested only
with 1-loop graphs and only with the real part thereof), it {\it
almost} suffices to use the simplified version of the formalism due to
Dolan and Jackiw \cite{DJ} (what {\it almost} means is explained in
the Appendix A).  In Dolan--Jackiw formalism the thermal propagators
present the sum of two terms: the standard zero-temperature propagator
and the temperature-dependent term which reflects the presence of real
particles in the heat bath (alias, the {\it temperature
insertion}). In particular, the thermal propagator of massless scalar
is
\begin{eqnarray}
\label{DJprop}
D_T(p) =  \frac i{p^2 + i0} + 2\pi \delta (p^2) \frac 1{e^{\beta |p_0|} - 1}
\end{eqnarray}

The temperature-dependent contribution in the polarization operator in
Fig. \ref{scalarloop} is
\begin{eqnarray}
\label{mqu}
	{\lambda \over 2} \int \frac {d^3p}{(2\pi)^3 |{\bf p}|
(e^{\beta |{\bf p}|} - 1)} = \frac{\lambda T^2}{24}
\end{eqnarray}
which is the familiar temperature renormalization of the scalar mass.
(Here and in the following, we ignore the infinite ultraviolet mass
renormalization in the zero-temperature theory which we assume to be
cancelled by a proper counterterm.)  Let us try to calculate the same
graph in classical theory. It is ultraviolet divergent
\footnote{Up to now, we assumed that the tree-level mass of the scalar
field is zero, but the result (\ref{mcl}) holds also for the theory
with nonzero mass $M_0$ if $\mu \gg M_0$. For the one-loop estimate
(\ref{mcl}) to be stable with respect to higher-order corrections, the
characteristic momenta in the integral $|{\bf p}| \sim \mu$ should
also be much larger than the full quantum temperature-induced mass
(\ref{mqu}) : $\mu \gg \lambda^{1/2} T$.}.
\begin{eqnarray}
\label{mcl}
M_{cl}^2(T) = \frac{\lambda T}2 \int \frac{d^3p}{(2\pi)^3 |{\bf p}|^2}
= \lambda \frac{T\mu}{4\pi^2}
\end{eqnarray}
Note that the method of \cite{grigoriev}-\cite{bochkarev} where the
correlator (\ref{QQ}) was estimated by solving the classical equations
of motion with the given initial conditions which then were averaged
over by integrating over classical phase space with the weight
$\exp\{-\beta H^{cl}\}$ like in (\ref{Zcl}) is equivalent {\em in
perturbation theory} to calculating the classical graphs. In Sect.2,
we shall illustrate this explicitly for $\lambda \phi^4$ theory.

The way out of this difficulty suggested in \cite{ambjorn} was to
choose $\mu \sim T$. The particular proportionality coefficient was
fixed by the requirement that the classical and quantum expressions
for the energy density (the analogs of (\ref{Equ}) and (\ref{Ecl}))
coincide. Note, however, that choosing $\mu$ in such a way, we would
still have a mismatch between the classical (\ref{mcl}) and quantum
(\ref{mqu}) expressions for the mass renormalization. This mismatch is
a manifestation of a simple fact that the classical description is
justified only for the low momentum modes.  But, as we have seen, the
low-energy sector of the theory does not entirely decouple from the
high momentum modes.  They get entangled together.  Thus, the
procedure used in \cite{ambjorn} may be not reliable
\footnote{ This question is, however, not clear at present. We return
to the discussion of this issue in the last section. We want also to
emphasize that the violent power ultraviolet divergences in
$m^{cl}(T)$ and other parameters of the effective hight-$T$
Hamiltonian are spesific for 4-dimensional theories. In
two-dimensional case considered in the first Grigoriev and Rubakov
paper, the situation is much more benign.}.

Our main observation, however, is that low momentum modes and
high momentum modes {\em can} in principle be disentangled. The
correct procedure should be the following. Let us introduce an
intermediate scale $\mu$ such that $gT \ll \mu \ll T$ and treat the
modes with momenta $|\vec{p}| > \mu$ and $|\vec{p}| < \mu$ in a
different way.  The high momentum modes can be explicitly integrated
over leading to the effective hard thermal loop Hamiltonian for the
low momentum modes \cite{pisarski}-\cite{Nair}. As the low energy
bound for momenta circulating in the loop is not zero but is equal to
$\mu$, this effective Hamiltonian would involve also some counterterms
depending on the parameter $\mu$. With the effective Hamiltonian in
hand, one can return back to the original GR procedure, solve the
effective equations of motion, and average over initial conditions
with the weight $\exp\{-\beta H^{eff}({\rm phase ~space})\}$. The
ultraviolet 3-dim momentum cutoff should be chosen to be of order of
$\mu$. The final result will be $\mu$-independent
\footnote{We were able to construct such a $\mu$-independent algorithm
{\it explicitly} only for the scalar theory. In Ref.\ \cite{alford}
such an algorithm has been discussed for a scalar theory in (2+1)
dimensions. The implementation of this idea in the gauge theory meets
serious technical difficulties. We shall discuss it at length later.}.

We emphasize that the success of this modified procedure depends on two
circumstances:
\begin{enumerate}
\item For high momentum modes, the classical treatment is wrong, but these
modes can be taken into account {\em perturbatively}.
\item For low momentum modes, perturbation theory breaks down and the
 classical real time equations of motion are essentially
 non-perturbative.  They cannot be evaluated by weak coupling methods.
 But the classical approach is justified here, and one can do a
 numerical calculation in the GR spirit.
\end{enumerate}

Note that the method of separating hard and soft modes where the
former can be treated perturbatively while the latter cannot, but the
influence of hard modes on the soft mode physics is taken into
account, is a standard one in many physical problems. E.g., it is the
basis of the operator product expansion technique and sum rules in QCD
\cite{Shif}.

In this paper, we first restrict ourselves with an illustrative
analysis for $\lambda \phi^4$ theory. Later we outline the procedure
to be used for gauge theories.  Note that the situation is much more
complicated in gauge theories when one computes the real time
evolution of classical fields.  For computing the real time evolution
of the gauge fields, one obtains Feynman diagrams of the type shown in
Fig.\ \ref{polop}.  When these diagrams are computed, one obtains a
correction of the form
\begin{eqnarray}
	\alpha  T^2 F^{\mu \nu } (k^0/|\vec{k}|)
\end{eqnarray}
where $k^\mu$ is the external momentum.  This function has a
 nontrivial dependence on $k^0/|\vec{k}|$.  For $ |\vec{k}|\sim gT$,
 this contribution is as large as the free particle kinetic energy
 terms, $g^{\mu \nu} k^2 - k^\mu k^\nu$ .  One can show, however, that
 so long as $ k^0\ge gT$ and $|\vec{k}| \ge gT$ the contribution of
 higher-order loops in the two-point function and other irreducible
 vertices is still suppressed and perturbation theory works. The
 shorthand for all these vertices is called the Braaten-Pisarski
 effective Lagrangian \cite{pisarski}.

The main problem which distinguishes hot gauge theories compared to
the simple $\lambda \phi^4$ case is that this effective Lagrangian is
highly nonlocal. One cannot straightforwardly write the effective
theory in a Hamiltonian form and write differential equations of
motion for real time evolution of initial thermal gauge field
configurations. However, in the recent works
\cite{Blaizot93,Nair,Blaizot94} this theory {\em has} been written in
the local Hamiltonian form.  The price which one had to pay is the
introduction of additional variables the integration over which
restores the original nonlocalities. Still, the local Hamiltonian
description of hot gauge theories exists. Unfortunately, such a local
effective Hamiltonian exists only if performing the integration in the
hard thermal loops over all loop momenta (without infrared cutoff
$\mu$). As was already mentioned, we were not able to derive
$\mu$-dependent piece in the effective HTL Hamiltonian. Our hope is,
however, that the contribution of these unknown terms in the baryon
nonconservation rate in interest is suppressed.

The organization of this paper is as follows: In the second section,
we reconsider the formalism of Grigoriev and Rubakov, and show
explicitly how the ultraviolet cutoff dependence arises.  In the
Section 3 we use this formalism to compute the lowest order
contribution to the propagator in the $\lambda\phi^4$ theory. In the
fourth section we present a path integral formalism for computing real
time correlation functions by a method which in lowest order
approximation gives a result similar to that of Grigoriev and Rubakov.
The effects due to hard modes are discussed in Sect.4.  We argue that,
for most practical purposes, one can evaluate the contribution of high
momentum modes in a Gaussian approximation to the functional integral
representation for the real time correlation function.  We explicitly
demonstrate this for $\lambda\phi^4$ theory.  We show that this
results in a formalism similar to that of Grigoriev and Rubakov except
that one should use the effective hard thermal loop Lagrangian plus
some counterterms needed to render the three dimensional classical
equations finite.  In Sect.5, we discuss hot gauge theories and, in
particular, the kinetic equation approach and the Nair's effective
Hamiltonian which may be the basis for the correct analysis of the
problem of interest. We also discuss the kinematic limits in which our
computation is valid.  In the last section, we summarize our results.
We describe what will be needed to be done before realistic
computations for 3+1 dimensional gauge theories may be performed.

There are also two technical Appendices devoted to two unsuccessful
attempts to isolate $\mu$-dependent terms in the effective Hamiltonian
for hot Yang-Mills theory.  In Appendix A we isolate the leading
cutoff-dependent terms in the soft multigluon thermal vertices with
momentum infrared cutoff.  Unfortunately, such a cutoff breaks gauge
invariance. As a result, the $\mu$ - dependent terms in the effective
lagrangian are not gauge invariant, and the classical problem cannot
be consistently posed. In Appendix B we display the problems one
meets when trying to do the same with a lattice cutoff (here the
problems are of the opposite kind. The lattice regularization is
gauge-invariant, but to isolate the $\mu$-dependent pieces is very
difficult technically) and perform some illustrative calcultations for
hot scalar QED.

\section{Ultraviolet Problems}
\setcounter{equation}0 In this section, we elucidate why the procedure
of Ambjorn et al.\ doesn't properly take into account the high
momentum modes with $k\sim T$.  This is demonstrated for the case of a
scalar theory with the action
\begin{eqnarray}
	S =  \int d^4x \left\{ \frac 12
(\partial_\mu \phi (x))^2 - \frac 12 M_0^2 \phi^2 (x)
 - {\lambda \over {4!}} \phi^4 (x) \right\}
\end{eqnarray}
The object we want to compute is the real time correlation function
\begin{eqnarray}
\label{Ctkp}
	C(t,\vec{k},\vec{p}) =
	<\phi (t,\vec{k}) \phi (0,\vec{p}) >_T
\end{eqnarray}
where
\begin{eqnarray}
	\phi (t,\vec{k}) = \int d^3x e^{-i\vec{k}\vec{x}}
	\phi (t,\vec{x})
\end{eqnarray}
In general, one can consider correlation functions of composite
operators, but for the points we wish to make, it is sufficient to
consider correlation functions of the elementary boson field.

The procedure of Grigoriev and Rubakov is  equivalent to the
following:  Assume an initial condition which has been determined
by random selection from a classical thermal distribution, that is,
with weight
\begin{eqnarray}
	\exp\{-\beta H(\Pi^\ini ,  \phi^\ini )\}
\label{weight}
\end{eqnarray}
where $\Pi $ is the momentum conjugate to $\phi$.  If we know
$\Pi^\ini $ and $\phi^\ini $, we have the initial conditions needed
for solving the classical equations of motion.

The next part of the recipe is to solve the classical equations of
motion.  This then determines $\Pi(t,\vec{k}) $ and $\phi(t,\vec{k})$.
To compute the correlation function, one then has to average over the
initial conditions.

The equation of motion
\begin{eqnarray}
&&      \left( {{d^2} \over {dt^2}} + \vec{k}^2 + M_0^2\right)
 \phi (t,\vec{k})\nn \\ &&
= - {\lambda \over {3!}} \int {{d^3k_1} \over {(2\pi )^3}}
{{d^3k_2} \over {(2\pi )^3}}~
\phi(t,\vec{k}_1) \phi(t,\vec{k}_2)
\phi(t,\vec{k}-\vec{k}_1- \vec{k}_2)
\end{eqnarray}
is solved perturbatively:
\begin{eqnarray}
\phi(t,\vec{k}) = \phi_1(t,\vec{k}) + \phi_2(t,\vec{k}) + \cdots
\end{eqnarray}
$ \phi_1(t,\vec{k}) $ is a superposition of plane waves:
\begin{eqnarray}
	\phi_1(t,\vec{k}) = { 1\over{2 \omega_k} } \left( a(\vec{k})
	e^{-i(\omega_k t- \vec{k}\vec{x})}+ a^*(-\vec{k})
	e^{i(\omega_k t- \vec{k}\vec{x})} \right)
\end{eqnarray}
where $a,a^*$ are determined by the initial conditions
and
\begin{eqnarray}
	\omega_k = \sqrt{ \vec{k}^2 + M_\cl^2 }
\end{eqnarray}
The mass $M_\cl$ is related to the tree level mass by
\begin{eqnarray}
	M_\cl^2 = M_0^2 + \delta M_\cl^2
\end{eqnarray}
The mass counterterm $\delta M_\cl^2 $ will be determined below.
$\phi_2(t,\vec{k})$ is obtained by solving
\begin{eqnarray}
	\left( {{d^2} \over {dt^2}} + \vec{k}^2 + M_\cl^2\right)
 \phi_2 (t,\vec{k}) &=& - {\lambda \over {3!}} \int {{d^3k_1} \over
 {(2\pi )^3}} {{d^3k_2} \over {(2\pi )^3}} \phi_1(t,\vec{k}_1)
 \phi_1(t,\vec{k}_2) \nn \\ && \phi_1(t,\vec{k}-\vec{k}_1-\vec{k}_2){}
 + \delta M_\cl^2 \phi_1(t,\vec{k})
\label{phi2}
\end{eqnarray}
with the initial conditions
\begin{eqnarray}
 \phi_2(0,\vec{k}) = 0,\quad  { d\over {dt} }\phi_2(0,\vec{k}) =0
\end{eqnarray}
To evaluate the Green's function (\ref{Ctkp}), we have to average
 $\phi^\ini(\vec{p})[\phi_1(t,\vec{k}) +\phi_2(t,\vec{k})]$ over the
 initial condition with the weight \mref{weight}:
\begin{eqnarray}
	C_\cl (t,\vec{k},\vec{p}) &=& Z_\cl^{-1} \int [d\Pi^\ini]
	[d\phi^\ini] \exp\{-\beta H_0[\Pi^\ini,\phi^\ini]\}\nn\\ &&
	\Bigg(1+\int d^3x \Bigg( \frac\lambda{4!}
	\left(\phi^{\ini}\right)^4(\vec{x})
	 -\frac12\delta M_\cl^2\left(\phi^{\ini}\right)^2(\vec{x})
	\Bigg)\Bigg)\nn\\ && \phi^\ini(\vec{p})\Big(\phi_1(t,\vec{k})
	+\phi_2(t,\vec{k})\Big)
\end{eqnarray}
where
\begin{eqnarray}
	H_0[\Pi,\phi]=\int d^3x \Bigg\{ \frac12 \Pi^2(\vec{x}) +
	\frac12 \phi(\vec{x})(-\nabla^2 + M_\cl^2 )\phi(\vec{x})
	\Bigg\}
\end{eqnarray}
One obtains
\begin{eqnarray}
	C_\cl (t,\vec{k},\vec{p})&=&
	(2\pi)^3\delta(\vec{p}+\vec{k})
{ T\over{2\omega_\vec{p}^2 } }
	\Bigg\{ \cos(\omega_\vec{p}t)\nn \\
&&{}- { 1\over{2\omega_\vec{p}}^2 }
	\Bigg( \lambda T \int
       { {d^3k}\over{ 4\omega_\vec{k}^2 } }
	- \delta M_\cl^2\Bigg)
	\Big( \omega_\vec{p} t \sin(\omega_\vec{p}t) + \cos(\omega_\vec{p}t)
 \Big)  \Bigg\}
\label{gcl}\end{eqnarray}
The momentum integral is ultraviolet singular and is regularized by
the cutoff $\mu$:
\begin{eqnarray}
\int        { {d^3k}\over{ 4\omega^2_{\vec{k}} } }
	= \frac{\lambda}{8\pi^2}T\mu
+{\cal O} (\frac{M_\cl^2}{\mu^2})
\end{eqnarray}
The contribution proportional to $t$ on the r.h.s.\ of eq.\ \mref{gcl}
is due to a resonance term in the r.h.s.\ of eq.\ \mref{phi2} and has
to vanish.  This determines the mass counterterm (this self-consistent
procedure is just equivalent to solving the Dyson equation with the
polarization operator calculated by the tadpole graph in Fig.\
\ref{scalarloop} in the classical limit):
\begin{eqnarray}
\delta M_\cl^2 = \frac\lambda{4\pi^2} T\mu + {\cal O} (\frac{M_\cl^2}{\mu^2})
\end{eqnarray}
Taking the Fourier transform of $C_\cl$ with respect to $t$ we obtain
the classical Greens function in momentum space as
\begin{eqnarray}
G_\cl(p) = \frac{T}{|p^0|} 2\pi\delta(p^2-M^2_\cl)
\end{eqnarray}
This has to be compared with the quantum mechanical expression.  The
summation of the diagrams in Fig. \ref{scalarmass} yields the real
time propagator
\begin{eqnarray}
	G(p) = {i \over {p^2 - M^2 +i\epsilon}} +2\pi \delta (p^2 -
        M^2) {1 \over {e^{\beta |p^0|} - 1}} \label{gqu}
\end{eqnarray}
with
\begin{eqnarray}
        M^2 = M_0^2 + \frac\lambda{24} T^2
\end{eqnarray}
The low energy limit eq.\ \mref{gqu} is
\begin{eqnarray}
        G(p) \sim \frac{T}{|p^0|} 2\pi\delta(p^2-M^2)
\end{eqnarray}
(It is also the classical limit. When substituting $p_0 \rightarrow
\hbar \omega, {\bf p} \rightarrow \hbar {\bf k}$, we see that the
second term in Eq.(\ref{gqu}) involves an extra power of $\hbar$ in
the denominator compared to the first term when $\hbar \rightarrow
0$).

Ambjorn et al.\ have choosen the cutoff such that the energy densities
of the quantum mechanical (eq.\ \mref{Equ}) and the classical system
(eq.\ \mref{Ecl}) are equal.  With this choice, however, there is a
mismatch of the self energies corresponding to the mass counterterms.

\section{A Formal Solution}
\setcounter{equation}0

As was emphasized in the introduction, the quantities of interest
which we wish to compute are real time correlation functions
\begin{eqnarray}
  \label{corrO} C(t,\vec{x}) & = & \langle {\cal O} (t,\vec{x}) {\cal
	O} (0)\rangle_T \nonumber\\ & = & { {Tr\{ e^{-\beta H}
	e^{itH}{\cal O} (0,\vec{x}) e^{-itH} {\cal O} (0) \}} \over
	{Tr~ e^{-\beta H}}}\nonumber \\ & = & { {Tr\{ e^{itH}
	e^{-\beta H} {\cal O} (0,\vec{x}) e^{-itH} {\cal O} (0) \}}
	\over {Tr~ e^{-\beta H}}}
\end{eqnarray}
 where ${\cal O}[\Phi(t, \vec{x})]$ is a local operator.
This correlation function can be expressed as a path integral as
\begin{eqnarray}
	C(t,\vec{x} ) = \int_{\Phi \mid_{x^0 = -i\beta}
= \Phi \mid_{x^0 = 0} }
[d\Pi] [d\Phi ]~ {\cal O} (t,\vec{x}) {\cal O} (0) e^{iS}
\end{eqnarray}
In this equation, the time variable $t$ is a real Minkowski time.  The
fields are periodic over the time $x^0 = 0$ to time $x^0 = -i\beta $.
The action has been written in terms of fields and their conjugate
momenta in the form
\begin{eqnarray}
	S = \int_C dx^0\int  d^3x \:  \left(  \Pi(x)\partial_0 \Phi(x)
	-{\cal H}(\Phi,\nabla\Phi)\right)
\end{eqnarray}
where $ {\cal H} $ is the Hamiltonian density.  The contour $C$ of
 integration in time is shown in Figure \ref{contour}.  It begins with
 a real time evolution from $ x^0 = 0$ to the time $ x^0 = t$.  It
 then takes a Euclidean path from $ x^0 = t$ to $x^0 = t -i\beta$.
 Finally $x^0 $ evolves backwards in time from $x^0 = t-i\beta$ to
 $x^0 = -i\beta $.  The periodic boundary conditions for the trace are
 at the end points of the time contour.

This path integral representation yields the expression of Grigoriev
and Rubakov in the stationary phase approximation $\delta {\rm
Im}(iS)=0$.  This requires $\Pi$ and $\Phi$ to be constant along the
Euclidean interval of the contour, i.e. $\Pi(x^0,
\vec{x})=\Pi_\ini(\vec{x})$, $\Phi(x^0 ,\vec{x})=\Phi^\ini(\vec{x})$
and to be a solution of the classical equations of motion along the
real time pieces. Then the phase vanishes in region $C_{II}$, while
the phases of regions $C_I$ and $C_{III}$ cancel each other.  The
exponential $\exp(iS)$ becomes $\exp\{-\beta H(\Pi^\ini,\Phi^\ini)\}$
and the only remaining integrations are those over $\Pi^\ini$ and
$\Phi^\ini$, which are the initial values for the real time evolution
in regions $C_I$ and $C_{III}$.  Then the correlation function is
\begin{eqnarray}
	C_{cl}(t,\vec{x}) = \int [d\Pi^\ini][d\Phi^\ini] e^{-\beta
	H(\Pi^\ini,\Phi^\ini)} {\cal O}_{cl} (t,\vec{x}) {\cal O}_{cl}
	(0)
\end{eqnarray}
This treatment ignores quantum corrections completely. It can be
justified if the only relevant modes are those with momenta much
smaller than the temperature since then their occupation number is
large and they can be treated classically. Since the calculation of
Grigoriev and Rubakov is performed on a lattice, it involves a cutoff
$\mu\sim 1/a$, where $a$ is the lattice spacing. From the above
argument $\mu$ should be much smaller than the temperature.  Grigoriev
and Rubakov have choosen $\mu\sim T$. If their result is
$\mu$-independent, it is probably reliable. However, if it depends on
$\mu$, it is sensitive to modes with momenta of order $T$, for which
the classical treatment is not correct.

For the latter case we suggest the following procedure: One should
choose the cutoff $\mu$ such that
\begin{eqnarray}
gT\ll\mu\ll T
\end{eqnarray}
(This condition is written for gauge theories. For the $\lambda
\phi^4$ theory the characteristic temperature-induced mass $gT$ should
be substituted by $\lambda^{1/2}T$).  The fields are decomposed into
short wavelength modes and long wavelength modes:
\begin{eqnarray}
	\Pi=\Pi_S + \Pi_L, \qquad \Phi=\Phi_S + \Phi_L
\end{eqnarray}
where $\Pi_S$ and $\Phi_S$ contain the Fourier componets with
$|\vec{k}|>\mu$, and $\Pi_L$, $\Phi_L$ consist of those with
$|\vec{k}|<\mu$.  Then the $\Phi_S$ interact weakly among themselves
and they can be treated as free fields in the background of $\Phi_L$,
i.e. for these modes one can use the Gaussian approximation in the
path integral. On the other hand, the condition $ \mu \ll T$ ensures
that the modes with $|\vec{k}|<\mu$ can be treated classically.

Naively we could ignore the Gaussian fluctuations around a classical
solution because it is higher order in coupling and, by assumption,
the coupling is weak.  In this case, however, there are corrections of
order $g^2 T^2$ to the classical equations of motion and these
generate a huge correction for the small momenta we wish to consider.
These terms arise from momenta which are much larger than the typical
frequencies we wish to consider for our classical solution, and
therefore the Gaussian fluctuations are dominated by the ultraviolet.
The integration over these modes is equivalent to integrating out the
high momentum modes and generating an effective theory of the low
momentum modes.

Let us examine in more detail the reason why it is a good
approximation to only do the Gaussian corrections to the path integral
in the presence of the long wavelength modes.  Consider modes with a
typical momentum scale $\mu $.  In this range of momentum, the
occupation number of the modes, $T/\mu$, is large compared to 1, but
small compared to $1/g$.  When these modes interact among themselves,
the typical strength of interaction is of order $g T/\mu \ll 1$.
Thus, nonlinear interactions among hard modes can be neglected, and the latter
can be treated in the quadratic (Gaussian) approximation. In a diagrammatic
language, it suffices to keep track only of the graphs involving 1-loop
subgraphs with high internal momenta (the so called hard thermal loops).

We therefore seek to integrate out the short wavelength modes to
obtain an effectively classical theory for the long wavelength modes.
This gives
\begin{eqnarray}
\label{Cnew}
	C(t,\vec{x}) \simeq  \int [d\Pi_L][d\Phi_L]
	e^{i(S[\Pi_L,\Phi_L]+\delta S[\Pi_L,\Phi_L;\mu]) }
	{\cal O} [\Phi_L (t,\vec{x})] {\cal O}[\Phi_L  (0)]
\end{eqnarray}
where
\begin{eqnarray}
	 \label{seff} \lefteqn{ e^{i\delta S[\Pi_L,\Phi_L;\mu] } }&&
	\nonumber \\ &= & \int [d\Pi_S][d\Phi_S] \exp\left\{i \int
	dx^0\left( \int d^3 x \left(\Pi_S\partial_0\Phi_S
	-\frac12\Pi_S^2 -\frac{\delta H(\Pi_L,\Phi_L)
	}{\delta\Phi(x)}\Phi_S(x) \right) \right.  \right.\nonumber \\
	&& {}- \left.\left.  \frac12\int d^3x\int d^3y \frac{\delta^2
	H(\Pi_L,\Phi_L)
	}{\delta\Phi(x^0,\vec{x})\delta\Phi(x^0,\vec{y})}
	\Phi_S(x^0,\vec{x})\Phi_S(x^0,\vec{y})) \right) \right\}
\end{eqnarray}
The effective theory described by Eqs. (\ref{Cnew}, \ref{seff}) will be the
one for which the GR treament will be applied.

\section{Integrating Out the Hard Modes}
\setcounter{equation}0

We have seen in the last section, that the effect of the hard modes is
to generate an effective action for the long wavelength modes.
Furthermore, it is sufficient to integrate them out in the Gaussian
approximation.  On the sight, this still appears as a tremendously
difficult task.  The exponent on the r.h.s of Eq. \mref{seff} contains
terms linear in $\Phi_S$.  They can be nonvanishing when $\Phi_L$
contains modes with momenta smaller but close to $\mu$ and the fields
would have to be shifted, $\Phi_S\to\Phi_S+\delta\Phi_S$. Furthermore
one has to sum all one loop diagrams with external momenta
$|\vec{p}_i| <\mu$. The loop integration over $\vec{k}$ involves theta
functions of the type $\Theta(|\vec{k}+\vec{P}_i|-\mu)$, where the
$\vec{P}_i$ are linear combinations of the external momenta, for each
internal line.  However, these complications arise only for relatively
hard modes in $\Phi_L$ with $|\vec{p}_i| \sim \mu$.  In the previous
section we have argued, that these modes are only weakly interacting
and the corresponding contribution to $\delta S$ should be small.
\footnote{In other words, the exact propagators and vertices with
external momenta which are much larger than the characteristic
temperature-induced mass coincide with tree propagators and vertices
plus small corrections.}

Therefore we can restrict ourselves to external momenta $|\vec{p}_i|
\sim gT\ll\mu$.  In this approximation $\delta S[\Pi_L,\Phi_L;\mu]$ is
given by the sum of the one loop diagrams with a factor $\Phi_L(x)$
for each external line, where the loop momentum is restricted by
$|\vec{k}| >\mu $.  Then it is possible, even for a nonabelian gauge
theory, to extract the leading terms of all $n$-point functions. This
is similar to the analysis by Braaten and Pisarski \cite{pisarski},
who extracted the leading contribution proportional to $g^2T^2$. The
only difference is that in our case there is a lower limit $\mu$ for
the $|\vec{k}| $ integration, where $\vec{k}$ is the loop momentum.

For illustration, let us consider the simple case of the
$\lambda \phi^4$-theory with the action
\begin{eqnarray}
	S = \int_C dx^0\int d^3x \left\{ \pi(x)\partial_0\phi(x) -
	\frac12 \left[\pi^2(x)+(\nabla\phi(x))^2 + m^2 \phi^2 (x)
	\right] - {\lambda \over {4!}} \phi^4 (x) \right\}
\end{eqnarray}
Here the leading contribution in the effective action is given only by
 the diagram in Fig.\ \ref{scalarloop}.  First the integration over
 $\pi(x)$ is performed. Then the free propagator for the short
 wavelength field is
\begin{eqnarray}
	iD_S(x_1,x_2)=\int\frac{d^4k}{(2\pi)^4}\Theta(|\vec{k}|-\mu)\rho_0(k)
	e^{-ik\cdot(x_1-x_2)}(\Theta_C(x_1^0-x_2^0)+n(k_0))
\end{eqnarray}
where $n(k_0)$ denotes the Bose distribution function,
$\Theta_C(x_1^0-x_2^0)$ is the step function along the contour, and
the spectral density $\rho_0(k)$ is given by
\begin{eqnarray}
	\rho_0(k)=2\pi \frac1{2\omega_k}
	\left(\delta(k^0-\omega_k)-\delta(k^0+\omega_k)
	\right)
	\label{spectral}
\end{eqnarray}
with $\omega_k=(\vec{k}^2+m^2)^{1/2}$.
 We obtain
\begin{eqnarray}
	\delta S = -i\frac\lambda 2 \int_C dx^0\int d^3x \phi^2(x)D_S(x,x)
	\label{deltastpf}
\end{eqnarray}
The value of $D_S(x,x)$ is independent of $x^0$. In particular, it is
the same on all three pieces of the contour. Therefore
Eq. \mref{deltastpf} can be written as
\begin{eqnarray}
	\delta S = \int_C dx^0\int d^3x \left(-\frac12 \delta m^2
	 \phi^2(x)\right) \label{deltas2pf2}
\end{eqnarray}
This expression contains an ultraviolet singularity, which is removed
by a mass renormalization at zero temperature. Then, in the limit
$m^2\ll\mu^2$, one obtains
\begin{eqnarray}
	\label{delm2}
	\delta m^2=\frac{\lambda}{24}\left(T^2-\frac6{\pi^2}\mu T\right)
\end{eqnarray}
Taking into account Eq.  \mref{deltas2pf2} when calculating the
correlation function has two effects.  First, it modifies the equation
of motion for the real time evolution in regions $C_I$ and $C_{III}$
of the time contour.  Secondly, it changes the Hamiltonian on the
Euclidean piece $C_{II}$, which gives the thermal weight when taking
the average over the initial conditions.

For $\lambda \phi^4$ theory, there are two equivalent ways to impose
an ultraviolet cutoff in the classical problem. First, one can just
expand the field in a Fourier series
  \begin{eqnarray}
  \label{Fourier}
  \phi(\vec{x}, t) = \sum_{\vec{n}} \phi_{\vec{n}} (t) e^{2\pi i \vec{n}
\vec{x} /L},
\end{eqnarray}
write the classical equations of motion in terms of Fourier
harmonics $\phi_{\vec{n}} (t)$ and impose the momentum cutoff
$|\vec{k}| = 2\pi |\vec{n}|/L \leq \mu$.  Technically, it is more
convenient, however, to define the theory on a Hamiltonian lattice. As
we show in Appendix B, the ultraviolet divergences in the one-loop
lattice graph can be easily singled out, and the lattice analog of
Eq.(\ref{mcl}) is given by
\begin{eqnarray}
	\label{Mphilat} M_{cl. lat.}^2(T) = \frac{\lambda T a^2}4
	\int_{-\pi/a}^{\pi/a} \frac{dp_1 }{2\pi} \int_{-\pi/a}^{\pi/a}
	\frac{dp_2 }{2\pi}\int_{-\pi/a}^{\pi/a} \frac{dp_3 }{2\pi}
	\frac{1}{ 3 - \sum_{i=1}^3 \cos (ap_i) } = \nonumber \\
	\frac{\lambda T}{4(2\pi)^3 a} \int_{-\pi}^\pi
	dx\int_{-\pi}^\pi dy \int_{-\pi}^\pi dz \frac{1}{3 - \cos x -
	\cos y - \cos z} = \frac{\lambda T}{4a} \int_0^\infty d\alpha~
	e^{-3\alpha} I_0^3(\alpha)
\end{eqnarray}
where $a$ is the 3-dimensional lattice spacing and $I_0(\alpha)$ is
the exponentially rising variety of Bessel function. We see that the proper
choice of
$a$ is
\begin{eqnarray}
	\label{amn} a = \frac {\pi^2}{\mu} \int_0^\infty
	d\alpha~ e^{-3\alpha} I_0^3(\alpha)
\end{eqnarray}
Then the formulae (\ref{Mphilat}) and (\ref{mcl}) match each other and
the final result for the correlators (\ref{corrO}) is
cutoff-independent.

\section{Gauge theories}
\setcounter{equation}0

Let us discuss now whether it is possible, and if yes then how, to
 generalize the method worked out in the previous sections for
 $\lambda \phi^4$-theory to an SU($N$) gauge theory.
\footnote{We shall ignore for time being the Higgs fields and
fermions.  Possible effects due to their presence will be discussed at
the end of the paper. Note that the studying of the correlator
(\ref{QQ}) makes sense also in hot $QCD$ where it determines the rate
of processes with chirality violation.}  As has already been mentioned
in the Introduction, this is not so trivial since integrating out the
modes with $|\vec{k}|\sim T$ does not simply result in a mass
renormalization as in Eq.(\ref{mqu}). One rather obtains nonlocal
terms proportional to $g^2T^2$, which for the modes with $|\vec{k}|\le
gT$ are at least as large as the terms in the classical action (for
the modes with $|\vec{k}| \sim g^2T$ they are even much larger!).
Furthermore, there is a leading $g^2T^2$ contribution to all $n$-point
functions. These terms have been identified in Ref. \cite{pisarski}
and their generating functional is known as the hard thermal loop
effective action. Since the resulting equations of motion are
nonlocal, they are not appropriate for computing correlation functions
a la Grigoriev and Rubakov.

Fortunately, this hard thermal loop effective field theory {\em can}
be presented in a local form \cite{Blaizot93,Nair}. In Ref.\
\cite{Blaizot93} it was formulated as a set of kinetic equations for
the long wavelength gauge fields and the induced current $j$ of hard
particles moving in the slowly varying background:
\begin{eqnarray}
	\label{Maxwell}
	D_\mu F^{\mu\nu}&= & j^\nu \label{kinAa}\\
	j^\nu(x) &=& 2 g N\int\frac{d^3p}{(2\pi)^3}
	v^\nu  \delta n(\vec{p},x)
\end{eqnarray}

Here $D_\mu $ denotes the covariant derivative in the adjoint
representation, $\delta n \equiv \delta n^a t^a$ is the deviation of
the thermal distribution of the hard particles from the equilibrium
Bose distribution function $n_0(|\vec{p}|)$.  Furthermore, $v^\mu$ is
the velocity of massless particles with momentum $p$,
$v^\mu=(1,\vec{v})$, $\vec{v}=\vec{p}/|\vec{p}|$.  The kinetic
equation for $\delta n$ is
\footnote{ The equation system (\ref{Maxwell}--\ref{kin}) presents the
nonabelian analog of the well-known Vlasov equations derived
originally for the usual non-relativistic plasma \cite{LP}. Silin
generalized them for ultrarelativistic abelian plasma \cite{Silin}. In
Ref.\cite{supersound} the kinetic equation approach has been used to
study the properties of supersymmetric field theories at finite
temperature.}
\begin{eqnarray}
	\label{kin}
	v^\mu D_\mu \delta n(x,\vec{p}) = - g \vec{v}\cdot\vec{E}
	\frac{d n_0(|\vec{p}|)}{d |\vec{p}| }
	\label{kinNa}
\end{eqnarray}
$\delta n$ depends nontrivially only on $\vec{v}$ and one can
integrate eq. \mref{kinNa} over $|\vec{p}|$.  On the r.h.s. of
equation \mref{kinNa} one has the integral
\begin{eqnarray}
	\int_0^\infty d |{\bf p}| |{\bf p}|^2 \frac{dn_0(|{\bf
	p}|)}{d|{\bf p}|}=-\frac13 \pi^2 T^2 \label{pint}
\end{eqnarray}
Then the induced current can be written in terms of
\begin{eqnarray}
 	\label{wdef} w(x, \vec{v}) = \frac 3{g\pi^2 T^2} \int \delta
	 n(x, \vec{p}) |\vec{p}|^2 d|\vec{p}|
\end{eqnarray}
as
\begin{eqnarray}
	j^\nu(x) =2 m^2 (T)\int\frac{d\vec{v}}{4\pi} v^\nu w(x,\vec{v})
\label{current}
\end{eqnarray}
 where $m^2(T) = \frac 32 \omega_{pl}^2(T)$ , and $\omega_{pl}^2(T) =
g^2T^2 N/9$ is the plasmon frequency.  The function $w$ satisfies the
equation
\begin{eqnarray}
	v^\mu D_\mu w(x,\vec{v}) = \vec{v}\cdot\vec{E} (x)
	\label{kinNb}
\end{eqnarray}

By solving eq. \mref{kinNb} the induced current can be written as a
functional of the gauge fields.  Then the eq. \mref{kinAa} takes its
nonlocal form obtained from the hard thermal loop effective action.
The energy of the system of gauge fields and induced currents is given
by
\begin{eqnarray}
	\label{Hamiltonian} H=\int d^3 x {\rm Tr} \left\{
	\vec{E}\cdot\vec{E} +\vec{B}\cdot\vec{B} +
	2m^2(T)\int\frac{d\vec{v}}{4\pi}
	w(x,\vec{v})w(x,\vec{v})\right\}
\end{eqnarray}

An alternative local formulation has been given in Ref. \cite{Nair} by
introducing auxillary fields $G(x, {\bf v})$, which are SU($N$)
matrices satisfying the constraint $G(x, -{\bf v}) = G^\dagger (x,{\bf
v})$.  In the gauge $A^a_0 = 0$ with $D_\pm= \partial_0 \pm {\bf
v}\cdot {\bf D}$, the Hamilton equations of motion are:
\begin{eqnarray}
  	\label{eqNair} E^a_i & = & \partial_0 A^a_i \nonumber \\
	\partial_0 E^a_i + \epsilon_{ijk}( D_j B_k)^a & = & \int
	d\vec{v}~ v_i (J^a_+ - J^a_-) \nonumber \\ ( D_- J_+)^a & = &
	- \frac {m^2(T)}{8\pi} ~E^a_i v_i
\end{eqnarray}
where
\begin{eqnarray}
 	 \label{Jpm} J^a_+ & = & \frac {m^2(T)}{2\pi} {\rm Tr} \{ it^a
	( D_+ G)G^{-1}\} \nonumber \\ J^a_- & = & - \frac
	{m^2(T)}{2\pi} {\rm Tr} \{ it^a G^{-1}( D_- G)\}
\end{eqnarray}
The Gauss law constraint
\begin{eqnarray}
	\label{Gauss}
	(\vec{D} \vec{E})^a + \int d\vec{v} (J^a_+ + J^a_-) \ = \ 0
\end{eqnarray}
should be imposed.  The two local formulations of the hard thermal
loop effective theory given by eqs. (\ref{kinAa},\ref{current},
\ref{kinNb}) and eqs. (\ref{eqNair} - \ref{Gauss}) have been shown to
be equivalent in Ref. \cite{Blaizot94}.

An important remark is in order here. The one-loop graphs provide the
dominant contribution to the effective Lagrangian when the external
momenta are of order $gT$. When the external momenta become as small
as $g^2T$ (and we presumably need that to solve the problem), the
contribution of certain higher-order graphs to the effective vertices
is of the same order as the contribution of hard thermal loops of
Figs.\ \ref{polop}, \ref{htl} \cite{Lebedev}.

Really, consider the two-loop graph for the gluon polarization
operator in the kinematic region where the external momentum $q$ is of
order $g^2T$ (see Fig.\ \ref{ladder}).  The extra loop brings about
the small factor $g^2$, but one can show that this loop involves the
"resounding denominators": the propagators with momenta $p$ and $p+q$
bring about the large factor
\begin{eqnarray}
T^4 \delta(p^2) \frac 1{(p+q)^2} \sim \frac {T^2}{pq} \sim \frac
1{g^2} \end{eqnarray} which compensates the smallness $\sim g^2$ so
that the contribution of the graph in Fig.\ \ref{ladder} is
unsuppressed. (An accurate analysis shows that such a large
contribution appears only if taking into account hard thermal loops in
the internal gluon propagator.)

In fact, all higher-loop {\it ladder} graphs contribute on the same
footing.  To find out the gluon polarization operator, the 3-point,
4-point etc.  vertices at small momenta $\sim g^2T$, one has to resum
all these ladders which is a formidable task (in Ref.\cite{Lebedev},
such a resummation has been done for the photon polarization operator
in abelian theory).

Fortunately, however, we do not need to do that. The matter is that
the integral for the graph in Fig.\ \ref{ladder} is saturated by the
region where the momentum of the rung of the ladder is soft:
$|\vec{k}| \sim gT$ \cite{Lebedev}.  In our procedure, the integration
should be done with the cutoff $|\vec{k}| > \mu \gg gT$ for {\it all}
loop momenta. In that case, the contribution of the graph in Fig.\
\ref{ladder} and all its higher-loop counterparts is suppressed again
and can be safely ignored. Thus, we are left with the standard
one-loop graphs in Figs.\ \ref{polop}, \ref{htl} \ with loop momenta
being restricted to be greater than the separation scale $\mu$.

 Our suggestion is to solve numerically the equation system
(\ref{eqNair}-\ref{Gauss}) rather than just equations of motion of the
free Yang-Mills Hamiltonian. This would provide a proper account of
the potentially dangerous high-momentum contributions. However, if one
just uses the effective Hamiltonian (\ref{Hamiltonian}) obtained after
integration over {\it all} momenta in the hard thermal loops, the
results would, generally, depend on the ultraviolet cutoff which has
to be imposed. If one wants to construct a cutoff-independent
algorithm, one should add to the Hamiltonian (\ref{Hamiltonian})
$\mu$- dependent counterterms so that the final result would not
depend on $\mu$. However, to find out these counterterms explicitly is
much more difficult task here than in the toy $\lambda \phi^4$ model.

For example, a simple-minded momentum ultraviolet cutoff is not
suitable as it breaks gauge invariance. If trying to include the
cutoff-dependent terms in the effective Hamiltonian (qualitatively,
they modify the thermal mass $m^2(T) \sim g^2T^2$ by the terms $\sim
g^2\mu T$), such a Hamiltonian would not commute with the Gauss law
constraint and the classical problem just cannot be posed. See
Appendix A for the detailed discussion.

Also, we do not know how to do it with a gauge-invariant lattice
ultraviolet cutoff. First, it is not so trivial to construct a lattice
version of the Nair system (\ref{eqNair}-\ref{Gauss}). But, even
assuming it might be done, the major problem is that the lattice
regularization breaks rotational invariance. As a result, the
counterterms are also not rotationally invariant (see Appendix B for
an illustrative calculation in scalar QED), their structure is
complicated, and we are not able to write down their general form.

Thus, the only practical recipe we can suggest is to take the
effective equations of motion for the full HTL Hamiltonian
(\ref{Hamiltonian}) and solve them numerically in the GR spirit with
an ultraviolet cutoff $\mu \gg gT$.  After averaging over initial
conditions, we would get the correlator (\ref{QQ}) .Though the result
would generally depend on the cutoff $\mu$, we still hope that such a
dependence would actually be weak. The arguments for this case and the
corresponding discussion will be presented in the last section.

\section{Discussion}
\setcounter{equation}0

But first let us discuss again the most important question: whether
the procedure suggested here is really necessary, or perhaps the
original procedure of Ref.\cite{ambjorn} is still correct and the
numerical calculations for the tree-level Hamiltonian (without
integrating over high momentum  modes) provide the results for the
correlator (\ref{QQ}) which do not really depend on ultraviolet
cutoff.

At present, we do not know the answer to this question, and all we can
do is to provide some arguments both {\em pro} and {\em contra}
regarding the Ambjorn et al.\ case. Let us start with the {\em pro}
arguments.

They come mainly from the observation that the physics associated with
the baryon number non-conservation in electroweak theory at ultrahigh
temperatures $T \gg T_c$ (and the physics of chirality
non-conservation in hot QCD) is related to the momentum region $\sim
g^2T$ and is essentially non-perturbative. The estimate (\ref{DelB})
for the rate of such non-conservation depends only on the magnetic
screening scale $\sim g^2T$ and not on the scale $gT$. It is
conceivable then that {\it all} relevant momenta in the loops for the
correlator (\ref{QQ}) are of order $g^2T$. In that case, the results
of the classical GR calculation are not sensitive to the ultraviolet
region and just do not depend on the cutoff $\mu$.

Essentially the same can be argued in a slightly different way. We  have
seen that the
dependence on $\mu$  generally comes together with the
dependence on the plasmon mass
scale $\sim gT$ [see Eqs. (\ref{delm2}, \ref{mTmu})].
 But the latter is probably not relevant because the
characteristic momentum scale, where the baryon number nonconservation
occurs, is of order $g^2T$ which is much less than $m(T) \sim gT$. If
so, the term $\sim \mu$ in $m^2(T, \mu)$ is also not relevant. And
that means that the $\mu$ - dependent pieces in the correlator
(\ref{QQ}) calculated in a classical procedure are also absent.  {\it
If} this guess is true, the results for the correlator (\ref{QQ})
calculated with the tree Lagrangian and the results obtained in our
procedure with the Braaten-Pisarski-Nair Lagrangian should just
coincide in spite of that the theories look completely different. The
huge effects due to effective mass and vertices renormalization should
all cancel out in the end.\footnote{There are two known examples where
the cancellation of the effects related to hard thermal loops occurs
in a physically observable quantity. One is the axial anomaly
\cite{anom}. Individual graphs for the divergence of anomalous
triangle are drastically changed if taking into account the
renormalization of Green's function and/or vertices in the heat bath,
but these effects completely cancel out in the sum of all graphs (
that just follows from the fact that the anomaly $\partial_\mu j^5_\mu
\sim {\rm Tr} (G \tilde{G})$ is an operator identity which does not
depend on whether one averages it over vacuum state or over a thermal
ensemble).  A similar cancellation of , at first sight, large effects
due to so called {\em anomalous damping} occurs in the low-frequency
electromagnetic polarization operator in quark-gluon plasma
\cite{Lebedev}.}  In this case, all the effective HTL Lagrangian stuff
may play the role of ``(ultra-)violet hands on an enamel wall'' \cite{Brusov} -
beautiful but irrelevant for physics.

However, there is also an argument {\em contra}. It comes from the
analysis of baryon nonconservation rate in the temperature region
below the phase transition point. In this region, the quasiclassical
approximation works and an analytical calculation is possible
\cite{arnold}. When $T_c - T \gg gT_c$,
  \begin{eqnarray}
  \label{sph}
\Gamma(\Delta B \neq 0) \propto \exp \left \{ - \frac {m_W(T)} {g^2 T}
\right \}
  \end{eqnarray}
where
  \begin{eqnarray}
  \label{mW}
m_W^2(T) = m_W^2(0) (1 - T^2/T_c^2)
  \end{eqnarray}
is the temperature-dependent $W$-mass. (At finite temperature, the
notion of mass is not uniquely defined. The mass (\ref{mW}) is defined
as a curvature of the effective potential, not as a pole of the
propagator which has a completely different behavior \cite{EWplasmons}
).

Thus, the mass renormalization due to a high frequency thermal loop is
essential when evaluating $\Gamma(\Delta B \neq 0)$ at $T \leq T_c$.
Perhaps, these effects are also essential in the region $T \gg T_c$.
In this case, the effective HTL Lagrangian is not a luxury but a
neccessary tool and one {\it should} solve the equations of motion
with the Hamiltonian (\ref{Hamiltonian}) to determine the infrared
asymptotics of the correlator (\ref{QQ}).

Up to now we neglected the effects due to Higgs and fermion
fields. The Higgs field is bosonic and can be treated in the same way
as the gauge field: one has first to calculate hard thermal loops to
determine the effective Lagrangian for the system including the gauge
and the scalar fields and to write it down in a local form by
introducing when necessary extra dynamic variables.  One should then
solve the equations of motion with this effective Lagrangian for soft
modes.

In fact, Higgs fields have been taken into account in the numerical
simulations in \cite{ambjorn}. It was found that their inclusion
practically does not affect the numerical results. That may be thought
of as an argument supporting the conjecture that only magnetic gauge
modes with momenta of order $g^2T$ are relevant for the problem.

Fermion fields play a different role because they cannot be treated
classically. One has to include them however in the loops when deriving
the effective Lagrangian for the gauge-Higgs system. If the naive GR
calculation is ultraviolet-sensitive, also the terms in the Lagrangian
due to fermion loops are important. If not - they are not.

Our suggestion to people who may wish to determine numerically the
behaviour of the correlator (\ref{QQ}) at very high temperatures (the
calculations in \cite{ambjorn} have been done in the region $T \sim
T_c$) is the following: First, forget about the Higgs field and the
fermion fields and do the calculations for a pure Yang-Mills system
with the ultraviolet cutoff $\mu \gg gT$ (but it may be chosen much
less than $T$ which is a considerable relief for a computer) and look
whether the results depend on the cutoff.  If they do not, the result
is probably reliable.  However, if they do, one has to redo the
calculations with the Braaten-Pisarski-Nair effective Lagrangian in
the way suggested in the paper.  If the results obtained with this
modified algorithm would not depend on the ultraviolet cutoff, one can
be sure that they are the answer.

To find the baryon number nonconservation rate in the hot electoweak
theory, one has first to derive the analog of Braaten-Pisarski
Lagrangian with account of Higgs fields and fermions which is yet to
be done.

Finally, let us discuss why one can hope that the modified procedure
would lead to results which practically do not depend on the cutoff
$\mu$.  Essentially, it is due to the already mentioned fact that
$\mu$-dependence in the GR procedure means the relevance of high
momenta in the loops in which case it is natural to expect that the
rate would depend not only on the scale $\sim g^2T$ , but also on the
plasmon mass scale $\sim gT$ which enters the HTL Hamiltonian
(\ref{Hamiltonian}). And that would mean that the estimate
(\ref{DelB}) is actually wrong (the heuristic arguments of
\cite{arnold} are physically appealing but they do not have the rank
of a rigorous proof). In this case, however, one need not bother so
much with carefullly subtracting the $\mu$-dependent counterterms
because they affect results only slightly ($g^2\mu T \ \ll \ g^2
T^2$).

For sure, this argumentation is not rigourous and the possibility
remains that the estimate (\ref{DelB}) is still correct and the scale
$\sim gT$ is irrelevant {\it but} the classical GR calculation {\it
is} sensitive to the ultraviolet. In this case one cannot expect that
the results obtained with the Hamiltonian (\ref{Hamiltonian}) are
$\mu$-independent, and the only way to solve the problem is to
accurately isolate $\mu$-dependent counterterms in the effective
Lagrangian with a gauge-independent (lattice) regularization
procedure.  We have seen that it is a very difficult task, and one
should summon the forces to attack this problem only if more fortunate
options --- i) GR procedure is not sensitive to the ultraviolet and
gives the correct answer and ii) Our modified procedure with the
Hamiltonian (\ref{Hamiltonian}) is not sensitive to the ultraviolet and
gives the correct answer --- are ruled out.

\section{Acknowledgments}

We are indebted to J. Ambjorn , Yu.M. Makeenko, V. Rubakov, D. Schiff,
M.S.  Shaposhnikov, and N. Turok for illuminating
discussions. A.S. acknowledges warm hospitality during his stay at the
University of Minnesota when this work was started.  support under DOE
High Energy DE-AC02-83ER40105 and DOE Nuclear DE-FG02-87ER-40328.

\section*{Appendix A. Soft gluon vertices with momentum cutoff.}
\renewcommand{\theequation}{A.\arabic{equation}}
\setcounter{equation}0

  This Appendix is devoted to the calculation of the hard thermal
loops for multi-gluon vertices with an infrared cutoff for internal
loop momentum $|\vec{p}| \ \geq \ \mu$ ($gT \ \ll \ \mu \ \ll \
T$). Although, as we shall see, the effective Hamiltonian generated by
these vertices breaks down gauge invariance and cannot be used to
solve the problem, it is important to understand where the problem
is. Also, this calculation may be interesting from a methodical
viewpoint. We consistently use the real time formalism which is
simpler and physically more transparent than the imaginary time
formalism standardly used.

  We start with estimating the contribution from the region $|\vec{p}|
\geq \mu $ in the gluon polarization operator. 3 relevant graphs are
depicted in Fig.\ \ref{polop}. We work in the simplified Dolan-Jackiw (DJ)
version \cite{DJ} of the real time technique which is equivalent to
the full-scale real time technique \cite{indus}-\cite{Landsman} in
this particular problem. The DJ expressions for the gluon and the
ghost propagators have the form (cf. Eq.(\ref{DJprop})
\begin{eqnarray}
\label{DJgauge}
D^{T,ab}_{\mu\nu}(p) =  - \delta^{ab} \left( g_{\mu\nu} -
\xi \frac { p_\mu p_\nu} {p^2} \right)
\left( \frac
i{p^2 + i0} + 2\pi \delta (p^2) \frac 1{e^{\beta |p_0|} - 1} \right)
\nonumber \\
D^{T,ab}_{ghost}(p) =  - \delta^{ab}
\left( \frac
i{p^2 + i0} + 2\pi \delta (p^2) \frac 1{e^{\beta |p_0|} - 1} \right)
    \end{eqnarray}
 where $\xi$ is the gauge fixing parameter.

 The 2-point vertex in the effective Lagrangian is given by the real
part of the temperature-dependent contribution in the polarization
operator which comes from the terms where the $\delta$-function
insertion in the propagator is taken into account only once.  In the
limit when the external momentum $q$ is considered to be much smaller
than the internal momentum $|\vec{p}| \sim \mu$, one can neglect
$q$-dependence in the vertices and the expression is considerably
simplified. Adding 3 graphs together, one gets after simple
transformations
\begin{eqnarray}
\label{Pimn}
\Pi_{\mu\nu}^{ab} (q) = 2g^2 N \delta^{ab} \int \frac {d^4p}{(2\pi)^3}
\frac {2p_\mu p_\nu -
g_{\mu \nu} p^2}
{(p+q)^2 - p^2} \left[ \delta[(p+q)^2] n_B (|p_0 + q_0|) -
\right. \nonumber \\     \left.
\delta (p^2) n_B (|p_0|) \right ] + O(q)
\end{eqnarray}
The imaginary time version of this formula can be found in
\cite{pisarski}. Note that the result is $\xi$-independent. This is
the notorious gauge-independence of hard thermal loops
\cite{pisarski}.

Next, we expand in $q$ the denominator $(p+q)^2 - p^2 = 2pq + O(q^2)$
and the expression in the square brackets in the integrand in Eq.
(\ref{Pimn}). In the latter, there are two terms coming from the
expansion of $n_B (|p_0 + q_0|)$ and of $\delta[(p+q)^2]$. When
integrating over $dp_0$, we have to take into account two roots of
$\delta$-function $p_0 = \pm |\vec{p}|$ which give, however, identical
contributions. Thus, it suffices to calculate the contribution of one
of the roots $p_0 = |\vec{p}|$.

The term involving $\delta'(p^2) = \frac 1{2p_0} \frac {\partial}
{\partial p_0} \delta (p^2)$ should be integrated by parts. The
derivative $\frac {\partial} {\partial p_0}$ acts upon $n_B(p_0)$ and
also on the structure
  \begin{eqnarray}
S_{\mu\nu} (p) = \ \frac {2p_\mu p_\nu - g_{\mu\nu}p^2}{p_0}
  \end{eqnarray}
It is not difficult to see that $\partial S_{\mu\nu} (p) /\partial
p_0$ is nonzero only for the {\it spatial} components $\mu = i, \nu =
j$.  In the remaining integral over $d|\vec{p}|$ the infrared cutoff
$|\vec{p}| \equiv \epsilon \geq \mu$ should be introduced. The loop
integrals for $\Pi_{00}^{ab}, \ \Pi_{i0}^{ab}, \ \Pi_{ij}^{ab}$ take
the form
\begin{eqnarray}
\label{allPi}
\Pi_{00}^{ab}(q) &\approx& 2 g^2 N \delta^{ab}
\frac1{(2\pi)^3}
\int d\vec{v} \frac {\vec{q} \vec{v}}{q_0 - \vec{q}
\vec{v}}
\int_{\mu}^{\infty} d \epsilon \: \epsilon^2 \: \frac{d
n_B(\epsilon)}{d\epsilon} \\
\Pi_{0i}^{ab}(q)&\approx&  2 g^2 N \delta^{ab}\frac{q^0q^i}{\vec{q}^2}
\int d\vec{v} \frac {\vec{q} \vec{v}}{q_0 - \vec{q}
\vec{v}}
\int_{\mu}^{\infty} d \epsilon \: \epsilon^2 \: \frac{d
n_B(\epsilon)}{d\epsilon} \\
\Pi_{ij}^{ab} (q) &\approx&  2 g^2 N \delta^{ab}\frac1{(2\pi)^3}
\int  d\vec{v} \left\{ v^iv^j \frac{\vec{q} \vec{v}}{q_0 - \vec{q}
\vec{v}}
 \int_{\mu}^{\infty}
 d \epsilon \: \epsilon^2 \: \frac{d n_B(\epsilon)}{d\epsilon} \right.
\nonumber \\ && {}
\left. -\frac23\delta_{ij}
\int_{\mu}^{\infty} d \epsilon \: \epsilon \: n_B(\epsilon)
\right\}
\end{eqnarray}
with $\vec{v} = \vec{p}/\epsilon$.  We see that the integrals for
$\Pi_{00}$ and $\Pi_{0i}$ factorize in the product of two independent
integrals: the energy integral and the angular integral. For
$\Pi_{ij}$, the expression involves two different energy integrals and
the factorization is not quite complete.

Without the infrared cutoff $\mu$ the two energy integrals would just
 coincide -- one would be obtained from the other by integrating by
 parts. When $\mu \neq 0$, this is not the case:
\begin{eqnarray}
\label{omder}
- \int_{\mu}^{\infty} d \epsilon \:
\epsilon^2 \: \frac{dn_B(\epsilon)}{d\epsilon }
= \frac{\pi^2}3 \left(T^2 -\frac3{\pi^2} \mu T\right)
\end{eqnarray}
and
\begin{eqnarray}
\label{om}
2 \int_{\mu}^{\infty} d \epsilon
 \: \epsilon \: n_B(\epsilon)
= \frac{\pi^2}3 \left(T^2 -\frac6{\pi^2} \mu T\right)
\end{eqnarray}
Since these integrals differ by a $\mu$- dependent term, the
polarization tensor in Eq. (\ref{allPi}) is not transverse (thereby
the gauge invariance is lost).  It is rather remarkable, however, that
this non-transverse piece does not actually depend on the gauge
parameter $\xi$ in the free thermal gluon propagators.

The appearance of non-transverse piece in the gluon polarization
operator with momentum cutoff is not a surprise. Just recall that the
usual zero-temperature photon polarization operator calculated with a
momentum ultraviolet cutoff involves a quadratically divergent photon
mass term.

What we are going to show next is that the troublesome mismatch
(\ref{omder}, \ref{om}) appears only for the 2-point function. For
$n\ge 3$ all $n$-point functions retain the same form as without
cutoff up to a universal renormalization of the coefficient.  Consider
first the 3-gluon vertex. In Feynman gauge there are two relevant
diagrams depicted in Fig.\ \ref{threepf} (The diagram containing a
four gluon vertex does not contribute at leading order in
$q$.). Let us calculate the real part of the effective vertex in
the Dolan-Jackiw technique. Note that here not only the terms
involving one thermal $\delta$-function but also the terms with the
product of 3 $\delta$-functions and 3 distribution functions
contribute. The contribution of these $\delta \otimes \delta \otimes
\delta$ term is huge $\sim g^3 T^3 / q^2_{char}$ where $q_{char}$ is a
characteristic external momentum. It is much greater than the tree
vertex in the region $q_{char} \sim gT$. Moreover, one can see that
the integral for this contribution is singular in the infrared and is
saturated by the internal momenta of order $q_{char}$. Thus, it is not
a hard thermal loop at all!

The resolution of this problem is known. The Dolan-Jackiw formalism
gives the so called $\Gamma^{111}$ component of the vertex in the
matrix real time formalism. What we need, however, is the {\it
retarded} vertex $\Gamma^R$ (the one which is obtained by analytic
continuation from the Euclidean region). $\Gamma^R$ presents the
combination of $\Gamma^{111}$ and also other components. The term
involving the product of 3 distribution functions cancels out in this
combination \cite{Schiff}.  One can show that the leading in $q$ terms
in $\Gamma^R$ {\it coincide} with those in $\Gamma^{111}$ which
involve only one temperature insertion
\footnote{Now we understand what {\it almost} in the discussion
preceding Eq.(\ref{DJprop}) meant.}.  As a result, the integral for
the temperature contribution to the 3-point vertex has the following
structure
\begin{eqnarray}
 \label{Gamma}
 \Gamma_{\mu\nu\lambda} \sim g^3 \int  \frac {d^4p}{(2\pi)^3} p_{1\mu}
p_{1\nu} p_{1\lambda} \left[
\frac {\delta (p_1^2)n_B(|p_{10}|)} {\delta_{21} \delta_{31}} +
\right.\nonumber \\
\left. \frac {\delta (p_2^2)n_B(|p_{20}|)} {\delta_{12} \delta_{32}} +
\frac {\delta (p_3^2)n_B(|p_{30}|)} {\delta_{13} \delta_{23}} \right ]
\end{eqnarray}
where $\delta_{kl} = p_k^2 - p_l^2$ and $p_k$, $k=1,2,3$ are the
internal momenta in the loop. $\delta_{kl}$ have the order $\sim
Tq_{char} \sim gT^2$ and are small. We have also neglected the
dependence on the external momenta in the tensor structure multiplying
the square bracket in (\ref{Gamma}).

The individual terms in the r.h.s.\ of Eq. (\ref{Gamma}) have the
order $\sim g^3 T^3 /q_{char}^2$ and are large. However, these large
contributions cancel in the total sum due to the algebraic
identity
\begin{eqnarray}
\label{S3}
S_3(1,2,3) = \frac 1 {\delta_{21} \delta_{31}} +
\frac 1 {\delta_{12} \delta_{32}} + \frac 1 {\delta_{13} \delta_{23}} =
0
\end{eqnarray}
As a result, the vertex has the normal HTL order $\sim g^3 T^2
/q_{char}$. To get this "non-leading" term, one has to expand the
numerators of the fractions in Eq.(\ref{Gamma}) in the external
momenta with respect to, say, the momentum $p_1$. As earlier, there
will be the terms coming from the expansion of the distribution
functions $n_B(|p_{20}|)$ and $n_B(|p_{30}|)$ which are proportional
to external energies and also, potentially, the terms coming from the
expansion of $\delta$-functions. The terms $\sim \delta'(p_1^2)$ are
dangerous as, when integrating by parts, the derivative may act upon
the structure $\sim p_\mu p_\nu p_\lambda /p_0$ and provide the
contribution involving the integral (\ref{om}) rather than
(\ref{omder}). That could give rise to some extra structures in the
vertices and, as a result, in the effective equations of
motion\mref{effeom}.

Fortunately, it does not happen. Really, the terms $\sim \delta'(p^2)$
in (\ref{Gamma}) are multiplied by
\begin{eqnarray}
  - \frac 1 {\delta_{32}} - \frac 1 {\delta_{23}} = 0
\end{eqnarray}
Thus, {\it only} the terms involving the expansion of $n_B(|p_{20}|)$,
$n_B(|p_{30}|)$ and thereby proportional to external energies {\it
and} involving only the integral (\ref{omder}) appear.

Let us prove it in general (for a similar proof in imaginary time
framework see \cite{Efraty}). The proof is inductive. Suppose we
already know that, for the $n$-point vertex, the large individual
contributions $\sim g^n T^3 / (q_{char})^{n-1}$ coming from the graphs
where $\delta$-function term is inserted in a particular internal line
\footnote{As earlier, only the terms with a single temperature
insertion contribute in the retarded $(n+1)$ - point vertex. } cancel
in the total sum, and terms coming from the expansion of the
$\delta$-function also cancel out. Let us prove it for the
$(n+1)$-point vertex.

The relevant graphs are shown in Fig. \ref{htl}.  The integral has the
form
\begin{eqnarray}
 \label{Gamman}
 \Gamma_{\mu_1 \ldots \mu_{n+1}} \sim g^{n+1} \int \frac
{d^4p}{(2\pi)^3} p_{1\mu_1} \ldots p_{1\mu_{n+1}} \left[
\frac {\delta (p_1^2)n_B(|p_{10}|)} {\delta_{21} \ldots \delta_{n+1,1}}
+
 \right. \nonumber \\ \left.
\ldots + \
\frac {\delta (p_{n+1}^2)n_B(|p_{n+1,0}|)} {\delta_{1,n+1} \ldots
\delta_{n,n+1}}       \right]
\end{eqnarray}
 Let us prove that
\begin{eqnarray}
\label{Sn}
S_{n+1}(1,\ldots,n+1) =
\frac 1 {\delta_{21} \ldots \delta_{n+1,1}} +
\frac 1 {\delta_{12} \ldots \delta_{n+1,2}} +
\cdots \nonumber \\ + \frac 1 {\delta_{1,n+1} \ldots
\delta_{n,n+1}} = 0
\end{eqnarray}
To this end, present the second term in the r.h.s.\ of Eq.(\ref{Sn}) as
\begin{eqnarray}
\frac 1 {\delta_{12} \ldots \delta_{n+1,2}} =
\frac 1 {\delta_{12} \delta_{32} \ldots \delta_{n+1,1}}
+ \frac 1 {\delta_{1,n+1} \delta_{32} \ldots
\delta_{n+1,2}}
\end{eqnarray}
and do the same for the third, $\ldots$, and $n$-th term in the sum.
Then it is not difficult to see that
\begin{eqnarray}
S_{n+1}(1,\ldots, n+1) = \frac 1 {\delta_{n+1,1}}
\left [ S_{n}(1,\ldots, n) - S_{n}(2,\ldots, n+1) \right]
\end{eqnarray}
which is zero by the inductive assumption.

To get a nonzero result, one has to expand $\delta(p_k^2)
n_B(|p_{k0}|)$ in the integrand in (\ref{Gamman}) in the external
momenta. The term proportional to $\delta'(p_1^2)$ involves the factor
\begin{eqnarray}
-\frac 1 {\delta_{32} \ldots \delta_{n+1,2}} -
\ldots - \frac 1 {\delta_{2,n+1} \ldots
\delta_{n,n+1}} \equiv -S_n(2,\ldots,n+1) = 0
\end{eqnarray}
Thus, an additional non-transverse structure involving the difference
of the integrals (\ref{om}) and (\ref{omder}) appears {\it only} in
the gluon polarization operator. Unfortunately, is it already bad
enough to spoil the game.

Let us try to write down the effective equations of motion
corresponding to the $n$-point functions calculated above. By means of
the time contour formalism of Sect.\ 3 one can see that indeed the
retarded Greens functions enter the equations of motion ensuring
causality. Thus we obtain
\begin{eqnarray}
      D_\mu F^{\mu 0} (x) = j^0 (x)\label{constraint}
\end{eqnarray}
\begin{eqnarray}
      D_\mu F^{\mu i}(x) = j^i(x)  + \frac1{3\pi^2} g^2 N\mu T A^i(x)
  \label{effeom}
\end{eqnarray}
The current $j^\mu$ is given by eqs.\ \mref{current}, \mref{kinNb}
with the plasmon mass $m_{pl}(T)$ replaced by $m_{pl}(T,\mu)$, where
 \begin{eqnarray}
 \label{mTmu}
 m^2_{pl}(T,\mu)=\frac16 g^2N \left(T^2 - \frac3{\pi^2} \mu T \right)
 \end{eqnarray}
The extra term $\propto A^i$ in eq.\ \mref{effeom} appears due to the
mismatch (\ref{omder})), (\ref{om}) discussed above.  Due to this term
the equations of motion \mref{effeom} are not consistent with Gauss'
constraint \mref{constraint}. This means that choosing initial
conditions satisfying the constraint (\ref{Gauss}), one cannot assure
that this constraint will be fulfilled also at later times, and the
classical problem cannot be consistently posed.

\section*{Appendix B. Thermal loops on a Hamiltonian lattice.}
\renewcommand{\theequation}{B.\arabic{equation}}
\setcounter{equation}0

We present here some illustrative calculations which display the way
the cutoff-dependence arises in the classical GR procedure with the
lattice ultraviolet regularization. Unfortunately, the results of this
study are negative --- for the gauge theory in interest we were not
able to extract explicitly the cutoff-dependent terms which is
necessary to construct an explicit lattice-based cutoff-independent
algorithm. But it is important to understand where the problem is. It
is conceivable that after considerable future efforts such an
algorithm would be eventually constructed.

Consider first the $\lambda \phi^4$ theory. The lattice Lagrangian
\footnote{It is more convenient to calculate graphs in the Lagrangian
formulation, but one should remember that our lattice is 3-dimensional
and the time is real and continuous } is
\begin{eqnarray}
   \label{Hphilat}
 {\cal L}_{lat} = \frac {a^3}{2} \sum_{\vec{n}} \dot{\phi}_{\vec{n}}^2
- \frac {a}2 \sum_{\vec{n}; i=1,2,3}(\phi_{\vec{n}+\vec{e}_i} -
\phi_{\vec{n}})^2 - \frac \lambda {4!} a^3 \sum_{\vec{n}}
\phi_{\vec{n}}^4
\end{eqnarray}
 where $\vec{e}_1 = (1,0,0)$ etc.  To calculate graphs, we have to go
over into the momentum representation (see \cite{Creutz} for
details). Actually, the only thing we need to know here is the
modified dispersive law
\begin{eqnarray}
  \label{displat}
p^2 \equiv p_0^2 - \frac 2{a^2} \sum_{i=1}^3 [1 - \cos (ap_i)]  = 0
\end{eqnarray}
 The 3-momenta $p_i$ range within the limits
\begin{eqnarray}
   \label{range}
-\pi/a \leq p_i \leq \pi/a
\end{eqnarray}
Let us calculate the graph in Fig.\ \ref{scalarloop} on the lattice in
the real time formalism.  The real time scalar lattice propagator has
the form (cf. Eq.(\ref{DJprop}) )
\begin{eqnarray}
  \label{DJlat}
  D_T(p) =  \frac i{ p^2
 + i0} + 2\pi \delta ( p^2)
\frac 1{e^{\beta |p_0|} - 1}
\end{eqnarray}
   In the limit $aT \gg 1$, we can, as earlier, take the classical
limit $T/|p_0|$ of the Bose distribution function after which the
thermal contribution to the scalar mass acquires the form
\begin{eqnarray}
  \label{MApplat}
M_{cl. lat.}^2(T) = \frac{\lambda T a^2}4 \int_{-\pi/a}^{\pi/a}
\int_{-\pi/a}^{\pi/a}\int_{-\pi/a}^{\pi/a}
 \frac{dp_1 dp_2 dp_3}{(2\pi)^3 [3 - \sum_{i=1}^3 \cos (ap_i)] }
\end{eqnarray}
With the help of the identities
\begin{eqnarray}
  \label{ident}
 \frac 1A = \int_0^\infty d\alpha e^{-\alpha A}, ~~~~\frac 1{2\pi}
\int_{-\pi}^\pi dx e^{\alpha \cos x} = I_0(\alpha),
\end{eqnarray}
 the integral can be reduced to a one-dimensional form as spelled out
in Eq.  (\ref{Mphilat}).

Let us go over now to the gauge theories. For illustrative purposes, we
restrict ourselves with the case of scalar QED. The part of the lattice
Lagrangian involving scalar fields is
\begin{eqnarray}
   \label{HQEDlat}
 {\cal L} = a^3 \sum_{\vec{n}} |{\cal D}_0 {\phi}_{\vec{n}}|^2
+ a \sum_{\vec{n}; i=1,2,3}
(\phi_{\vec{n}+\vec{e}_i}^*
U_{\vec{n}+\vec{e}_i,\ \vec{n}} \phi_{\vec{n}}
+ \nonumber \\
\phi_{\vec{n}+\vec{e}_i}
U_{\vec{n}+\vec{e}_i,\ \vec{n}}^* \phi_{\vec{n}}^*
- \phi_{\vec{n}+\vec{e}_i}^* \phi_{\vec{n}+\vec{e}_i}
- \phi_{\vec{n}}^* \phi_{\vec{n}})
\end{eqnarray}
where ${\cal D}_0 = \partial_0 - ie A_{0 \vec{n}}$ and
$U_{\vec{n}+\vec{e}_i, \vec{n} }$ are complex fields
living on the links
of the lattice . (They have the meaning of parallel transporters:
$U_{\vec{n}+\vec{e}_i, \vec{n} } \rightarrow \exp\{ieaA_i(\vec{x})\}$
in the continuum limit.)

The temperature contribution to the photon polarization operator is
determined by the graphs in Fig.\ \ref{scalarqed}. In the continuum
limit, it is given by the same integral as in Eq.(\ref{Pimn}) . Let us
impose the lattice ultraviolet cutoff and assume that $aT \gg 1 \gg
aq$ where $q$ is the external momentum.

Consider first the component $\Pi_{00}$ of the polarization
operator. It is a little bit simpler than others because the vertices
are the same as in the continuum theory. For $\Pi_{00}$ the only
modification brought about by the lattice are the limited range of
integration (\ref{range}) over spatial momenta and the modified
dispersive law (\ref{displat}). As a result, the integral for
$\Pi_{00}$ reads
\begin{eqnarray}
   \label{Pint}
\Pi_{00} = 2e^2T \int dp_0 \int_{-\pi/a}^{\pi/a} \frac{dp_1}{2\pi}
\int_{-\pi/a}^{\pi/a}
\frac{dp_2}{2\pi} \int_{-\pi/a}^{\pi/a} \frac{dp_3}{2\pi} \frac {2p_0^2
 - p^2}{(p+q)^2 - p^2} \nonumber \\
\left[ \frac {\delta [(p+q)^2]}{|p_0 + q_0| } - \frac {\delta
(p^2)}{|p_0|} \right]
\end{eqnarray}
(where as earlier we neglected small $q$ compared to $p \sim 1/a$
whenever possible). Expanding $1/|p_0 + q_0|$ and $\delta[(p+q)^2]$ in
$q$, we get for the term $\propto 1/a$
\footnote{
cf. the expression (\ref{allPi}) with the momentum cutoff.}
\begin{eqnarray}
  \label{P00lat}
\Pi_{00} = - \frac {2e^2T}a \int_{-\pi}^\pi \int_{-\pi}^\pi
\int_{-\pi}^\pi
\frac {d^3s}{(2\pi)^3} \: \frac {\sum_i \: q_i \sin s_i}{q_0 - \frac
1{\omega_s } \sum_i \: q_i \sin s_i} \: \frac 1{\omega_s^3}
\end{eqnarray}
where $s_i = ap_i$ and
\[
\omega_s^2 = 2 \sum_i (1 - \cos s_i)
\]
And now we see that $\Pi_{00}$ is not the function of only $q_0$ and
$\vec{q}^2 = q_1^2 + q_2^2 + q_3^2$ as is the case in the continuum
theory, but depends on each component of $\vec{q}$ separately.  It is
more or less obvious from the form of the integral. But if a reader
cherishes a hope that it might not be true, he is welcome to expand
the expression (\ref{P00lat}) over $\vec{q}/|q_0|$ up to the fourth
order and to be explicitly convinced that on top of the rotationally
invariant structure $\sim (q_1^2 + q_2^2 + q_3^2)^2$ also the
structure $\sim q_1^2 q_2^2 + q_2^2 q_3^2 + q_1^2 q_3^2$ pops out with
a non-zero coefficient.

For completeness, we present here also the results for the other
components of $\Pi_{\mu \nu}$ :
\begin{eqnarray} \label{P0ilat}
\Pi_{0i} = - \frac {2e^2T}a q_0 \int_{-\pi}^\pi \int_{-\pi}^\pi
\int_{-\pi}^\pi \frac {d^3s}{(2\pi)^3} \: \frac { \sin s_i}{q_0 -
\frac 1{\omega_s } \sum_i \: q_i \sin s_i} \: \frac 1{\omega_s^3}
\end{eqnarray}
\begin{eqnarray}
\label{Pijlat}
\Pi_{ij} = - \frac {2e^2T}a q_0 \int_{-\pi}^\pi \int_{-\pi}^\pi
\int_{-\pi}^\pi
\frac {d^3s}{(2\pi)^3} \: \frac { \sin s_i \sin s_j}{q_0 - \frac
1{\omega_s } \sum_k \: q_k \sin s_k} \: \frac 1{\omega_s^4}
\end{eqnarray}
As the lattice regularization does not break gauge invariance,
$\Pi_{\mu \nu}$ of Eqs.(\ref{P00lat} - \ref{Pijlat}) is transverse:
$q_\mu \Pi_{\mu\nu} = 0$.

But, as the rotational invariance is broken, the cutoff-dependent
counterterm in the 2-point vertex of the effective HTL Lagrangian is
not presented as the sum of just two standard (transverse and
longitudinal) tensor structures as was the case in the effective
theory without cutoff. The same refers to counterterms for
multiple-gluon vertices which we have not calculated.

\newpage

{\large\bf Figure Captions} \\

\begin{enumerate}\renewcommand{\labelenumi}{Figure \arabic{enumi}}

\fig{scalarloop} The contribution to the propagator which yields the
effective temperature dependent mass in $\lambda \phi^4$ theory.

\fig{polop} The polarization contribution to the vector boson
propagator at finite temperature. The dashed lines represent ghost
propagators.

\fig{scalarmass} The sum of self energy insertions which yield the
propagator in $\lambda \phi^4$ theory.

\fig{contour}  The contour of the time integral in the path integral
representation for the correlation function.

\fig{ladder} Two-loop ladder graph contributing to the effective
action for very soft momenta $q \sim g^2T$.

\fig{threepf} Leading contributions to the effective three gluon
vertex in Feynman gauge. The dashed lines represent ghost propagators.

\fig{htl} Hard thermal loop for a $(n+1)$ - gluon vertex. The
vertical dashes stand for the temperature ($\delta$-function) insertions.
There are also similar graphs with internal ghost loop.

\fig{scalarqed} One-loop contributions to the photon polarization
operator in scalar QED.

\end{enumerate}

\newpage

\begin{center}
\begin{picture}(300,50)(0,0)

\CArc(150,20)(20,0,360)\Line(100,0)(200,0)

\end{picture}\\ \vspace{2cm}
Figure \ref{scalarloop}\\
\end{center}

\begin{center}
\begin{picture}(400,200)(0,0)

\PhotonArc(100,120)(20,0,360){1.5}{15}
\Photon(50,98)(150,98){1.5}{12}
\put(65,80){$k$}

\PhotonArc(313,120)(20,0,360){1.5}{15}
\Photon(250,120)(292,120){1.5}{5}
\Photon(332.5,120)(383,120){1.5}{5}
\put(270,102){$k$}

\DashCArc(100,-5)(20,0,360){3}
\Photon(50,-5)(80,-5){1.5}{5}
\Photon(120,-5)(150,-5){1.5}{5}
\put(65,-23){$k$}

\end{picture}\\ \vspace{2cm}

Figure \ref{polop}

\end{center}

\newpage

\begin{center}\begin{picture}(400,50)(0,0)

\CArc(25,20)(20,0,360)
\Line(-5,0)(55,0)

\put(102,20){$+$}
\CArc(180,20)(20,0,360)\CArc(240,20)(20,0,360)\Line(150,0)(270,0)
\put(317,20){$+\quad\cdots$}

\end{picture}\\ \vspace{2cm}

Figure \ref{scalarmass}

\end{center}

\begin{center}
\begin{picture}(300,300)(0,0)

\Line(0,75)(300,75)
\Line(100,-25)(100,175)

\Line(105,70)(270,70)
\Line(270,70)(270,0)
\Line(105,0)(270,0)

\CArc(279,154)(15,0,360)
\put(275,150){$x^0$}

\put(268,85){$t$}
\put(75,-3){$-i\beta$}
\put(90,65){$0$}
\end{picture}\\ \vspace{2cm}

Figure \ref{contour}

\end{center}

\newpage

\begin{center}\begin{picture}(300,100)(0,0)

\PhotonArc(150,-5)(20,0,360){1.5}{15}
\Photon(100,-5)(130,-5){1.5}{5}
\Photon(170,-5)(200,-5){1.5}{5}
\Photon(150,-26)(150,14){1.5}{5}
\put(105,-15){$q$}
\put(135,-32){$p$}
\put(120,22){$p+q$}
\put(155,-8){$k$}

\end{picture}\\ \vspace{2cm}

Figure \ref{ladder}

\end{center}

\begin{center}

\begin{picture}(300,100)(0,15)

\PhotonArc(75,20)(20,0,360){1.5}{15}
\Photon(95,20)(125,20){1.5}{5}
\Photon(65,37.32)(50,63.3){1.5}{5}
\Photon(65,2.68)(50,-23.3){1.5}{5}
\put(112,10){$q_1$}
\put(62,55){$q_2$}
\put(43,-12){$q_3$}
\put(85,45){$p_1$}
\put(40,20){$p_2$}
\put(83,-8){$p_3$}

\end{picture}

\begin{picture}(300,-300)(-130,0)

\DashCArc(75,20)(20,0,360){3}
\Photon(95,20)(125,20){1.5}{5}
\Photon(65,37.32)(50,63.3){1.5}{5}
\Photon(65,2.68)(50,-23.3){1.5}{5}
\put(112,10){$q_1$}
\put(62,55){$q_2$}
\put(43,-12){$q_3$}
\put(85,45){$p_1$}
\put(40,20){$p_2$}
\put(83,-8){$p_3$}

\end{picture}
\\
\vspace{2cm}

Figure \ref{threepf}

\end{center}

\newpage

\begin{center}
\begin{picture}(200,100)(130,0)

\PhotonArc(150,20)(20,0,360){1.5}{15}
\Photon(100,20)(130,20){1.5}{5}
\Photon(170,20)(200,20){1.5}{5}
\Photon(160,37.32)(175,63.3){1.5}{5}
\Photon(140,37.32)(125,63.3){1.5}{5}
\Photon(140,2.68)(125,-23.3){1.5}{5}
\Vertex(167.5,-10.3){1}
\Vertex(176.8,-2.5){1}
\Vertex(156,-14.5){1}
\Line(150,35)(150,45)
\put(224,20){$+$}

\end{picture}

\begin{picture}(200,100)(-30,-100)

\PhotonArc(150,20)(20,0,360){1.5}{15}
\Photon(100,20)(130,20){1.5}{5}
\Photon(170,20)(200,20){1.5}{5}
\Photon(160,37.32)(175,63.3){1.5}{5}
\Photon(140,37.32)(125,63.3){1.5}{5}
\Photon(140,2.68)(125,-23.3){1.5}{5}
\Vertex(167.5,-10.3){1}
\Vertex(176.8,-2.5){1}
\Vertex(156,-14.5){1}
\Line(163,27.5)(171.65,32.5)
\put(215,20){$+\cdots$}

\end{picture}
\\ \vspace{-2cm}

Figure \ref{htl}

\end{center}

\begin{center}
\begin{picture}(300,100)(0,0)

\CArc(75,0)(20,0,360)
\Photon(25,-22)(125,-22){1.5}{15}
\put(35,-32){$q$}
\put(75,30){$p$}

\CArc(238,0)(20,0,360)
\Photon(175,0)(217,0){1.5}{6}
\Photon(258,0)(308,0){1.5}{6}
\put(185,-10){$q$}
\put(238,30){$p$}

\end{picture}\\ \vspace{2cm}

Figure \ref{scalarqed}

\end{center}

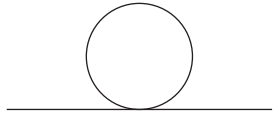
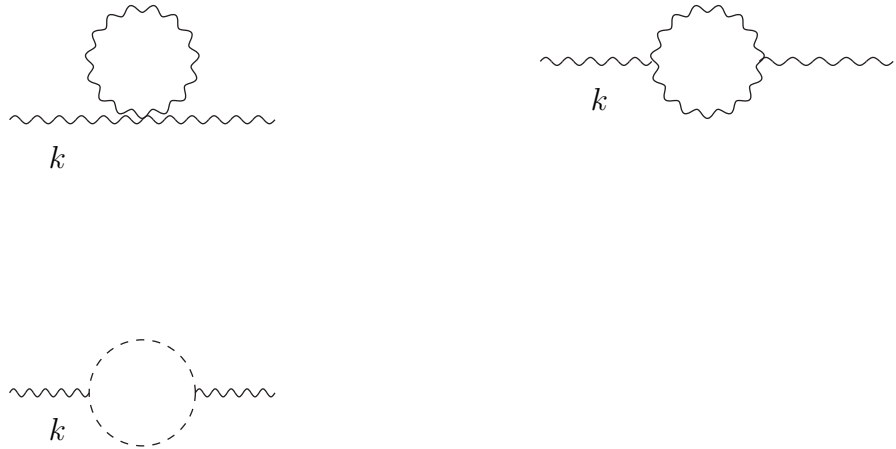
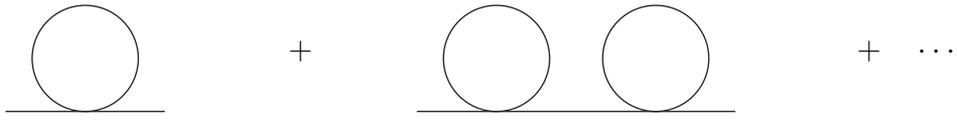
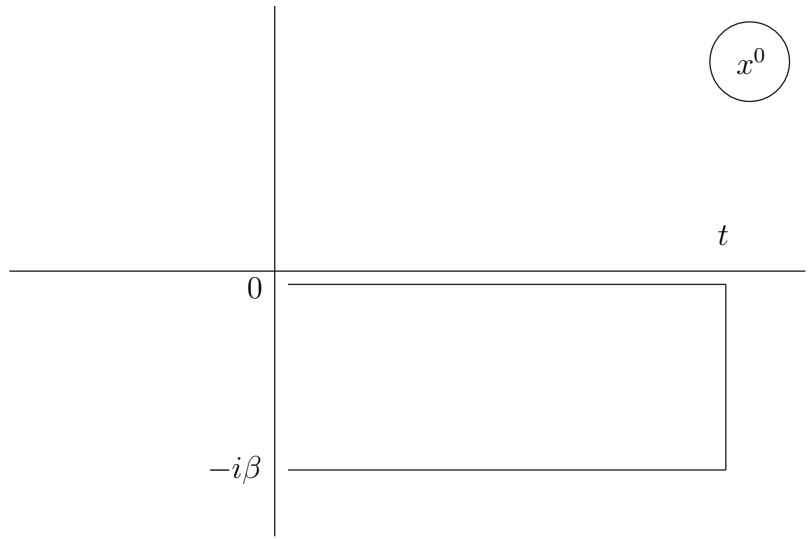
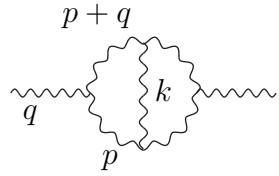
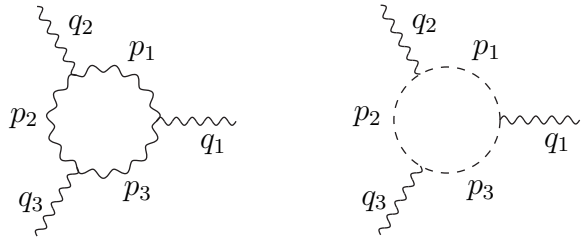
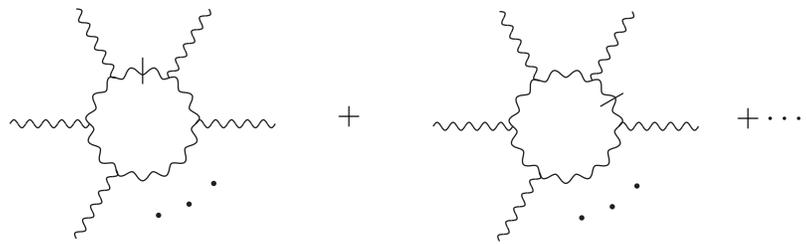
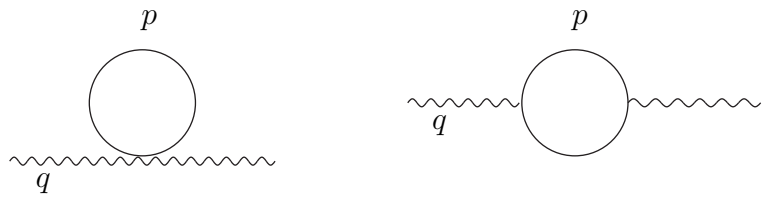
\end{document}